\documentclass[12pt]{spieman}  %>>> 12pt font mandatory; use this for US letter paper 
\usepackage{amsmath} 
\usepackage{amssymb}
\usepackage[]{graphicx}
\usepackage{setspace}
\usepackage{tocloft}

\title{Simulation of  nanostructure-based and
ultra-thin film solar cell devices beyond the classical picture} 

\author{Urs Aeberhard}

\affiliation{IEK-5 Photovoltaik, Forschungszentrum J\"ulich, D-52425 J\"ulich, Germany}

\cftpagenumbersoff{figure}
\cftpagenumbersoff{table} 
\begin{document} 
\maketitle 

%%%%%%%%%%%%%%%%%%%%%%%%%%%%%%%%%%%%%%%%%%%%%%%%%%%%%%%%%%%%% 
\begin{abstract}
In this paper, an optoelectronic device simulation framework valid for arbitrary spatial variation of electronic
potentials and optical modes, and for transport regimes ranging from ballistic to diffusive, is used
to study non-local photon absorption, photocurrent generation and carrier extraction in ultra-thin
film and nanostructure-based solar cell devices at the radiative limit. Among the effects that are
revealed by the microscopic approach and which are inaccessible to macroscopic models is the impact
of structure, doping or bias induced nanoscale potential variations on the local photogeneration
rate and the photocarrier transport regime.
\end{abstract}

%>>>> Include a list of up to six keywords after the abstract

%>>>> Include contact information for corresponding author
{\noindent \footnotesize{\bf Urs Aeberhard, IEK-5 Photovoltaik, Forschungszentrum J\"ulich, D-52425 J\"ulich, Germany ;
Tel: +49 2461 61 26 15; Fax: +49 2461 61 37 35 ; E-mail:
\linkable{u.aeberhard@fz-juelich.de} }
%%%%%%%%%%%%%%%%%%%%%%%%%%%%%%%%%%%%%%%%%%%%%%%%%%%%%%%%%%%%%

\begin{spacing}{2}   % use double spacing for rest of manuscript

%%%%%%%%%%%%%%%%%%%%%%%%%%%%%%%%%%%%%%%%%%%%%%%%%%%%%%%%%%%%%
\section{Introduction}
\label{sect:intro}  % \label{} allows reference to this section

With the introduction of novel light-trapping schemes reaching beyond the ray-optics
limit, high-efficiency solar cells with an active absorber thickness of only a fraction of the
typical irradiation wavelength are becoming interesting alternatives to expensive wafer-based
architectures. In the case where these ultra-thin film solar cells are made of high-mobility
semiconductors, the classical picture of local charge carrier generation and diffusive transport
in thermalized distributions may no longer be appropriate, especially in the presence of strong
doping-induced internal fields. The same applies to a wide range of nanostructure-based
photovoltaic device components, such as quantum well and quantum dot structures or highly doped tunnel junctions,
where the local electronic structure deviates strongly from the flat band bulk picture
conventionally assumed in photovoltaic device simulations.

In this situation, microscopic theories on the quantum kinetic level allow for the step beyond the local and macroscopic
description by enabling the consideration of arbitrary potential variations and general
nonequilibrium carrier distributions at the nanoscale. For the specific case of photovoltaic devices, where optical
transitions and charge carrier transport are equally important, such a theory has recently been
formulated based on the non-equilibrium Green's function formalism (NEGF) and has been applied
successfully to the investigation of a number of nanostructure based solar cell architectures \cite{ae:prb_08,ae:nrl_11,ae:oqel_12,ae:pccp_12,ae:prb87_13}.

In the present study, the focus is on the impact of structure, doping or bias induced nanoscale
potential variations on the local photogeneration rate and the photocarrier transport regime, as
these features are not accessible in macroscopic approaches\cite{ae:spie14}.

\section{Theory framework}

% - NEGF formalism for PV
% - only equations for electrons
% - description of self-energies for coupling to photons and phonons
% - description of light propagation (classical)
% - recombination: radiative only, isotropic photon DOS (-> no light trapping), no photon recycling 
% 
% - verification via flat-band bulk case (absorption/photogeneration) -> reference PRB

The general NEGF theory of nanostructure-based solar cell devices is described in detail in Ref.~\citenum{ae:jcel_11}. Here, only the elements relevant for the current investigation will be
presented.  In consideration of the high computational cost associated with the NEGF
simulation approach, the current application of the formalism is restricted to the technologically 
most relevant class of thin-film devices containing planar nanostructures.

\subsection{Green's functions and self-energies for planar systems}
In the case of planar device architectures, the charge carriers are described by a slab
representation for the steady-state Green's functions (GF), 
\begin{align}
G(\mathbf{r},\mathbf{r}',E)=&\mathcal{A}^{-1}\sum_{\mathbf{k}_{\parallel}}G(\mathbf{k}_{\parallel},z,z',E)
e^{i\mathbf{k}_{\parallel}\cdot(\mathbf{r}_{\parallel}-\mathbf{r}'_{\parallel})},
\end{align}
where $\mathcal{A}$ is the slab cross section area.  The classical electromagnetic field is expressed via the vector potential
\begin{align}
\mathbf{A}(\mathbf{r},t)=&\mathcal{A}^{-\frac{1}{2}}\sum_{\mathbf{q}_{\parallel}}\mathbf{A}(\mathbf{q}_{\parallel},z,t)
e^{i\mathbf{q}_{\parallel}\cdot\mathbf{r}_{\parallel}}.
\end{align} 
The steady state charge carrier GF are obtained from the 
integro-differential equations
\begin{align}
G^{R}(\mathbf{k}_{\parallel},z,z',E)=&G_{0}^{R}(\mathbf{k}_{\parallel},z,z',E)+\int dz_{1}\int dz_{2}~G_{0}^{R}(\mathbf{k}_{\parallel},z,z_{1},E)\Sigma^{R}(\mathbf{k}_{\parallel},z_{1},z_{2},E)G^{R}(\mathbf{k}_{\parallel},z_{2},z',E),\label{eq:slabgf_dyson}\\
G^{\lessgtr}(\mathbf{k}_{\parallel},z,z',E)=&\int dz_{1}\int dz_{2}~G^{R}(\mathbf{k}_{\parallel},z,z_{1},E)\Sigma^{\lessgtr}(\mathbf{k}_{\parallel},z_{1},z_{2},E)
G^{A}(\mathbf{k}_{\parallel},z_{2},z',E),\label{eq:slabgf_keldysh}
\end{align}
with
\begin{align}
\big[E-\mathcal{H}_{0}(\mathbf{k}_{\parallel},z)\big]G_{0}^{R}(\mathbf{k}_{\parallel},z,z',E)&=\delta(z-z').
\end{align}
The non-interacting part $\hat{H}_{0}$ of the Hamiltonian contains the electronic structure (at
this stage a simple two band effective mass model) and the electrostatic mean-field potential from coupling to the
Poisson equation via the non-equilibrium carrier densities,
\begin{align}
n(z)=&\mathcal{A}^{-1}\sum_{\mathbf{k}_{\parallel}}\int
\frac{dE}{2\pi}\big[-i G_{cc}^{<}(\mathbf{k}_{\parallel},z,z,E)\big],\quad
p(z)=\mathcal{A}^{-1}\sum_{\mathbf{k}_{\parallel}}\int \frac{dE}{2\pi}\big[i G_{vv}^{>}({\mathbf k}_{\parallel},z,z,E)\big].
 \label{eq:steaddens}
\end{align}
The term
$\hat{H}_{e\gamma}=-\frac{e}{m_{0}}\hat{\mathbf{A}}(\mathbf{r,t})\cdot\hat{\mathbf{p}}$ is used to
describe the interaction of charge carriers with photons that is required for radiative generation
and recombination processes via corresponding non-local self-energies, e.g., for electrons in the conduction band, 
\begin{align}
\Sigma_{cc}^{<(gen)}(\mathbf{k}_{\parallel},z,z',E)=&\Big(\frac{e}{m_{0}}\Big)^{2}
\sum_{\mu}p_{cv}^{\mu}(z)p_{cv}^{\mu*}(z')\int dE_{\gamma}
G_{vv}^{<}(\mathbf{k}_{\parallel},z,z',E-E_{\gamma})\mathcal{A}^{-1}\sum_{\mathbf{q}_{\parallel}}A_{\mu}(\mathbf{q}_{\parallel},z,E_{\gamma})
A_{\mu}^{*}(\mathbf{q}_{\parallel},z',E_{\gamma}), \label{eq:se_coh}\\
\Sigma_{cc}^{>(rec)}(\mathbf{k}_{\parallel},z,z',E)=&\Big(\frac{e}{m_{0}}\Big)^{2}\sum_{\mu,\nu}p^{\mu}_{cv}(z)p_{cv}^{\nu*}(z')
\int dE_{\gamma} G_{vv}^{>}(\mathbf{k}_{\parallel},z,z',E-
E_{\gamma})\mathcal{A}^{-1}\sum_{\mathbf{q}_{\parallel}}i\hbar\mu_{0}\mathcal{D}^{>}_{\mu\nu}(\mathbf{q}_{\parallel},z,z',E_{\gamma})\\
\approx&\frac{n_{0}^3}{3\pi\hbar
c_{0}^3\varepsilon_{0}}\Big(\frac{e}{m_{0}}\Big)^{2}\sum_{\mu}p^{\mu}_{cv}(z)p_{cv}^{\mu*}(z') \int dE_{\gamma} G_{vv}^{>}(\mathbf{k}_{\parallel},z,z',E-
E_{\gamma}),
\end{align}
where the local approximation of the momentum averaged GF of free field photon modes,
\begin{align}
\mathcal{A}^{-1}\sum_{\mathbf{q}_{\parallel}}\mathcal{D}^{>}_{\mu\nu,0}(\mathbf{q}_{\parallel},z,z',E_{\gamma})&\approx-\frac{i
n_{0}^3E_{\gamma}}{3\pi\hbar c_{0}}\delta_{\mu\nu},\qquad (z\approx z'),
\end{align}
was used in the last line. This corresponds to emission into an optically homogeneous medium. The classical electromagnetic vector potential $\mathbf{A}$ in the multilayer device is obtained
from the conventional transfer matrix method (TMM). The local extinction coefficient used in
the TMM is related to the local absorption coefficient as obtained from the microscopic interband
polarization in terms of the charge carrier Green's functions \cite{ae:prb_11}. At this point, it should be noted that for
detailed balance to hold, both self-energies should be expressed in terms of the same photon GF, 
which can be obtained from an additional set of Dyson and Keldysh equations similar 
to \eqref{eq:slabgf_dyson} and \eqref{eq:slabgf_keldysh}\cite{ae:oqel_13}.

In addition to the electron-photon interaction, the coupling of
charge carriers to phonons also needs to be considered for the description of relaxation effects and
phonon-mediated transport processes. Here, it is included via the standard self-energy on the level
of the self-consistent Born approximation based on the non-interacting equilibrium Green's
functions of bulk phonon modes \cite{mahan:87}.

\subsection{Absorption, generation rate and photocurrent}
The local and spectral photogeneration rate $g$ at fixed photon energy ($\sim E_{\gamma}$), polarization ($\sim \mu$) and incident angle ($\sim
\mathbf{q}_{\parallel},~E_{\gamma}$), where $\mathbf{q}_{\parallel}$ is the transverse photon momentum, is related to the corresponding local absorption coefficient $\alpha$ and local photon flux $\Phi$ via
\begin{align}
g^{\mu}(\mathbf{q}_{\parallel},z,E_{\gamma})=\Phi_{\mu}(\mathbf{q}_{\parallel},z,E_{\gamma})	
\alpha_{\mu}(\mathbf{q}_{\parallel},z,E_{\gamma}).\label{eq:genrate_def}
\end{align}
In the NEGF formalism, the spectral photogeneration rate can be obtained from the expression for the local integral
radiative interband generation rate $\mathcal{G}$ in terms of electronic Green's functions and
self-energies \cite{ae:prb_11}, which for charge carriers in the conduction band reads
\begin{align}
\mathcal{G}_{c}(z)=&\mathcal{A}^{-1}\sum_{\mathbf{k}_{\parallel}}\int dz'
\int\frac{dE}{2\pi\hbar}\Sigma_{cc}^{<(gen)}(\mathbf{k}_{\parallel},z,z',E)
G_{cc}^{>}(\mathbf{k}_{\parallel},z',z,E)
\equiv\mathcal{A}^{-1}\sum_{\mu}\sum_{\mathbf{q}_{\parallel}}\int
dE_{\gamma} ~g^{\mu}_{c}(\mathbf{q}_{\parallel},z,E_{\gamma}).\label{eq:integral_genrate}
\end{align}
At the radiative limit, the short circuit current density $J_{sc}$
is directly given by the incident photon flux and the total absorptance of the slab,
\begin{align}
J_{sc}=\frac{e}{\mathcal{A}}\sum_{\mathbf{q}_{\parallel}}\int
dE_{\gamma}\boldsymbol{\Phi}(\mathbf{q}_{\parallel},z_{0},E_{\gamma})\cdot
\mathbf{a}(\mathbf{q}_{\parallel},z_{max},E_{\gamma}).\label{eq:jsc_abs}
\end{align}
On the other hand, the
recombination-free limit of the short circuit current derives from the quantities computed within the NEGF formalism as
follows\cite{ae:prb_11}:
\begin{align}
J_{sc}=&j_{c}(z_{max})-j_{c}(z_{0})=\int_{z_{0}}^{z_{max}} dz ~\partial_{z}j(z)\equiv
e\int_{z_{0}}^{z_{max}} dz~\mathcal{G}_{c}(z),\label{eq:jsc_gen}
\end{align}
where $j_{c}$ denotes electron current in the conduction band, which is given terms of the charge
carrier Green's functions via
\begin{align}
j_{c}(z)=\lim_{z'\rightarrow
z}\frac{e\hbar}{m_{0}}(\partial_{z}-\partial_{z'})\mathcal{A}^{-1}\sum_{\mathbf{k}_{\parallel}}
\int\frac{dE}{2\pi}G_{cc}^{<}(\mathbf{k}_{\parallel},z,z',E).
\end{align}
Together, Eqs.~\eqref{eq:integral_genrate}-\eqref{eq:jsc_gen} yield the following expression of the
absorptance in terms of the local generation spectrum:
\begin{align}
a_{\mu}(\mathbf{q}_{\parallel},z_{max},E_{\gamma})=&\Phi^{-1}_{\mu}(\mathbf{q}_{\parallel},z_{0},E_{\gamma})\int_{z_{0}}^{z_{max}}dz~
g^{\mu}(\mathbf{q}_{\parallel},z,E_{\gamma}).
\label{eq:absorpt_gen}
\end{align} 
Using the electron-photon self-energy
\eqref{eq:se_coh} in expression \eqref{eq:integral_genrate} for
$\mathcal{G}$, the local spectral photogeneration acquires the following form:
\begin{align}
g^{\mu}(\mathbf{q}_{\parallel},z,E_{\gamma})=&\frac{i}{\hbar\mu_{0}}
A_{\mu}(\mathbf{q}_{\parallel},z,E_{\gamma})\int dz'
A^{*}_{\mu}(\mathbf{q}_{\parallel},z',E_{\gamma})
\Pi_{\mu\mu}^{>}(\mathbf{q}_{\parallel},z',z,E_{\gamma}),\label{eq:rate_coh}
\end{align}
where $\Pi$ is the photon self-energy related to the non-equilibrium polarization function $\mathcal{P}$ and the momentum matrix elements $p_{cv}$,
\begin{align}
\Pi_{\mu\nu}^{>}(\mathbf{q}_{\parallel},z,z',E_{\gamma})=&-i\hbar\mu_{0}\Big(\frac{e}{m_{0}}\Big)^{2}p_{cv}^{\mu*}(z)
\mathcal{P}_{cv}^{>}(\mathbf{q}_{\parallel},z,z',E_{\gamma}) p_{cv}^{\nu}(z'), \label{eq:photse_out}
\end{align}
with the random-phase-approximation of the interband polarization function given in terms of the
charge carrier GFs as follows:
\begin{align}
 \mathcal{P}_{cv}^{>}(\mathbf{q}_{\parallel},z,z',E_{\gamma})=&\mathcal{A}^{-1}\sum_{\mathbf{k}_{\parallel}}\int
 \frac{dE}{2\pi\hbar}G_{cc}^{>}(\mathbf{k}_{\parallel},z,z',E)
 G_{vv}^{<}(\mathbf{k}_{\parallel} -\mathbf{q}_{\parallel},z',z,E-E_{\gamma}).\label{eq:polfun}
\end{align}
Using \eqref{eq:rate_coh} in \eqref{eq:absorpt_gen}, the final expression for the absorptance of a
slab of thickness $d=z_{max}-z_{0}$ acquires the form
\begin{align}
a_{\mu}(\mathbf{q}_{\parallel},z_{max},E_{\gamma})=&\frac{i}{\hbar\mu_{0}}\Phi^{-1}_{\mu}(\mathbf{q}_{\parallel},z_{0},E_{\gamma})\int_{z_{0}}^{z_{max}}dz\int_{z_{0}}^{z_{max}}dz'
A_{\mu}(\mathbf{q}_{\parallel},z,E_{\gamma})A^{*}_{\mu}(\mathbf{q}_{\parallel},z',E_{\gamma})\Pi_{\mu\mu}^{>}(\mathbf{q}_{\parallel},z',z,E_{\gamma}).
\end{align}
The local absorption coefficient $\alpha$ which is required to provide the extinction coefficient
\begin{align}
\kappa_{\mu}(\mathbf{q}_{\parallel},z,E_{\gamma})=\alpha_{\mu}(\mathbf{q}_{\parallel},z,E_{\gamma})
\cdot\frac{\hbar c_{0}}{2E_{\gamma}}
\end{align}
used in the TMM is formally defined by \eqref{eq:genrate_def}
in terms of the local values of photon flux and photogeneration. If the local variation of the
electromagnetic field is neglected, an expression can be found that contains solely the local
electronic properties,
\begin{align}
\alpha_{\mu}(\mathbf{q}_{\parallel},z,E_{\gamma})=&\frac{\hbar c_{0}}{2
n_{r}(\mathbf{q}_{\parallel},z,E_{\gamma})E_{\gamma}}\int dz' 
\mathrm{Re}\Big[i\Pi_{\mu\mu}^{>}(\mathbf{q}_{\parallel},z',z,E_{\gamma})\Big] 
\label{eq:locabscoef}.
\end{align}
 
\subsection{Recombination and radiative dark current}
In analogy to the photogeneration process, a local rate of radiative recombination can be expressed in terms of the
emission self-energy and carrier GF,
\begin{align}
\mathcal{R}_{c}(z)=&\mathcal{A}^{-1}\sum_{\mathbf{k}_{\parallel}}\int dz'
\int\frac{dE}{2\pi\hbar}\Sigma_{cc}^{>(rec)}(\mathbf{k}_{\parallel},z,z',E)
G_{cc}^{<}(\mathbf{k}_{\parallel},z',z,E)
\equiv\mathcal{A}^{-1}\sum_{\mu}\sum_{\mathbf{q}_{\parallel}}\int
dE_{\gamma} ~r^{\mu}_{c}(\mathbf{q}_{\parallel},z,E_{\gamma}),
\end{align}
which defines the transverse momentum average of the local and spectral radiative emission rate $\bar{r}$,
\begin{align}
\bar{r}^{\mu}_{c}(z,E_{\gamma})\equiv
\mathcal{A}^{-1}\sum_{\mathbf{q}_{\parallel}}r^{\mu}_{c}(\mathbf{q}_{\parallel},z,E_{\gamma})\approx&\frac{
n_{0}^3}{6\pi^2\hbar^2c_{0}} \int dz' \mathrm{Re}\Big[i\Pi_{\mu\mu}^{<}(\mathbf{0},z',z,E_{\gamma})\Big] ,
\end{align}
where the photon self-energy component $\Pi^{<}$ is given by \eqref{eq:photse_out} with ``$>$'' replaced by ``$<$''. The
radiative dark current is then obtained as the spatial integral of the local recombination rate, 
\begin{align}
J_{dark,rad}=e\int_{z_{0}}^{z_{max}} dz~\mathcal{R}_{c}(z)\label{eq:j_rec}.
\end{align}

\section{Numerical simulation results for showcase structures}
In the following, deviations from the bulk or flat-band picture are revealed by applying the above
formalism to specific structures encountered as components of advanced nanostructure based solar
cell architectures, such as ultra-thin films, superlattices and heterostructure tunnel junctions. In all cases, the electronic structure is described by a simple two band model of a direct semiconductor, using either the effective mass approximation (EMA) or an $sp_{z}$ tight-binding (TB) approach, with parameters given in Tab.~\ref{tab:par_ema} in the appendix (EMA) or in Ref.~\citenum{ae:prb87_13} (TB). Intraband relaxation and energy dissipation is considered via inclusion of inelastic coupling of charge carriers to a single polar optical phonon mode as well as elastic coupling to acoustic phonons, and the extinction coefficient used for the light propagation is computed according to the local approximation in Eq.~\eqref{eq:locabscoef}.

\subsection{Field dependent generation and recombination in ultra-thin film absorber solar cells}
 %\begin{itemize}
%   \item convergence of JDOS to bulk value at flat band and finite field
%	\item schematic representation of thin film cell
%	\item ldos and band profile at V=0 and V=Voc 
% \item dark JV for two lower doping densities $\rightarrow$ change in J0 and n$_id$ 
  
 %\end{itemize}

\begin{figure}[b]
\begin{minipage}{5.5cm} 
 \begin{center}
\includegraphics[height=5cm]{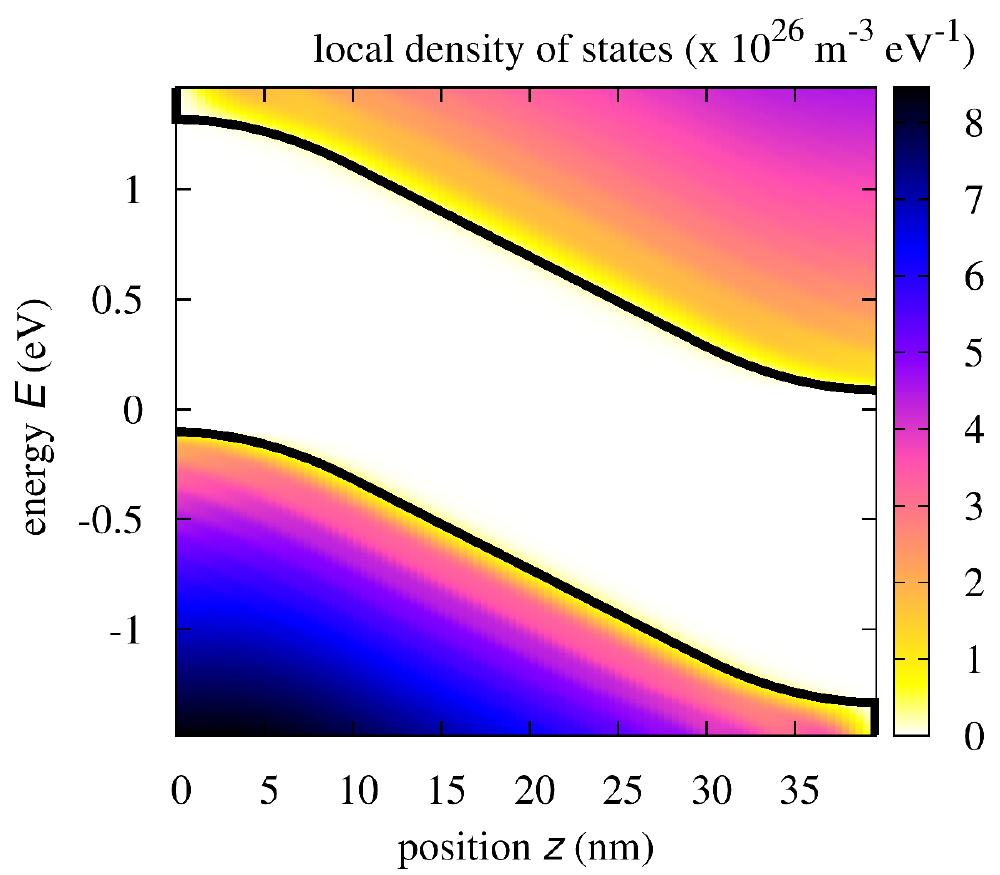}
 \caption{Band profile and local density of states of a 40 nm GaAs $pin$
 junction solar cell. The doped layers extend over 10 nm and provide a
 charge carrier density of 3$\times 10^{18}$ cm$^{-3}$.
 \label{fig:ldos_bulk_pin}}\vspace{1.2cm}
 \end{center}
 \end{minipage}\qquad
\begin{minipage}{10cm}
 \begin{center}
 \includegraphics[height=4.8cm]{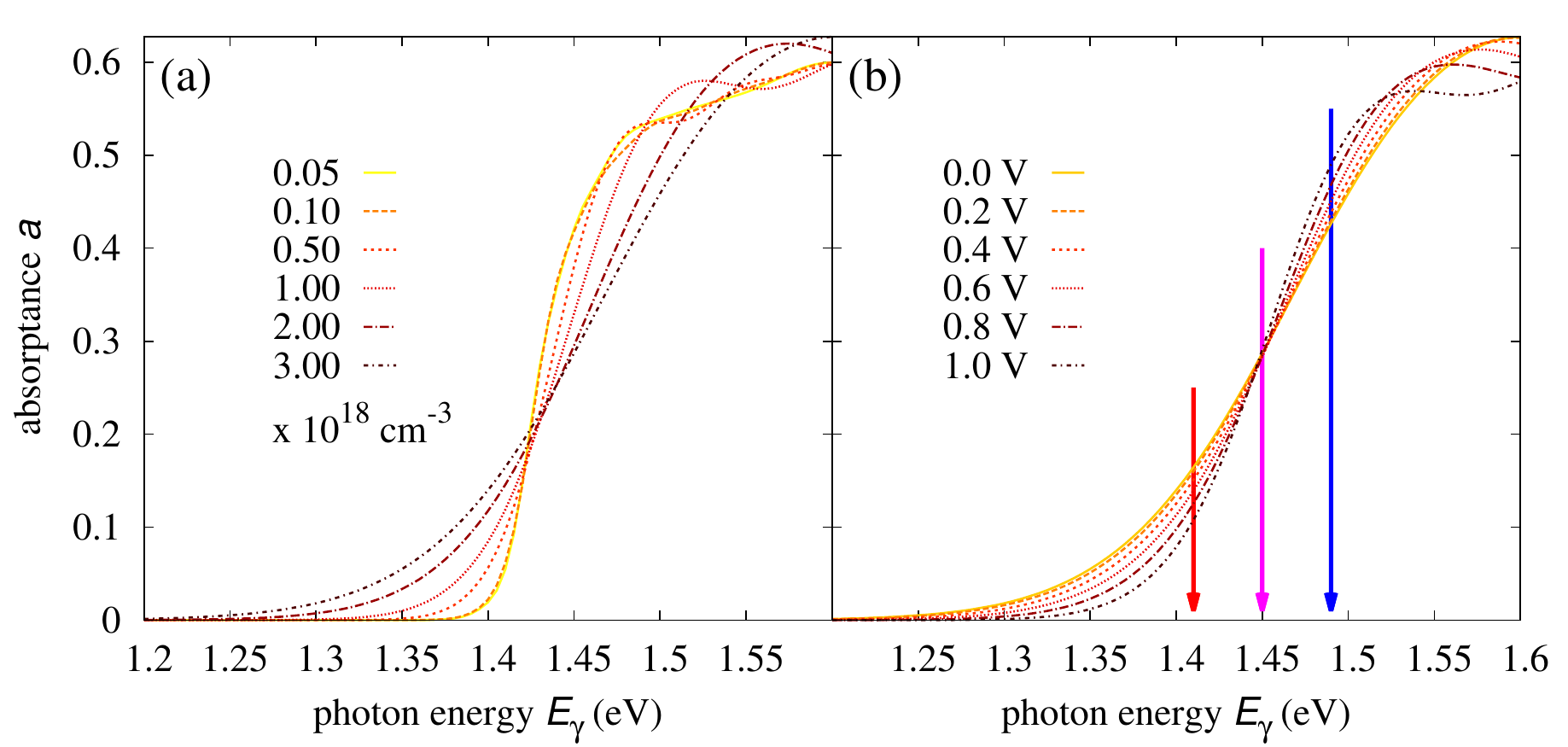}
 \caption{The strong built-in potential present in the ultra-thin junction
 leads to a Franz-Keldysh tailing of the band edge, inducing the absorption of sub-band gap energy
 photons. Following the strength of the built-in field, the tailing (a) increases with doping
 density and (b) decreases with forward bias. The effect on the
 absorbance depends on the photon energy: while the absorption is
 increased at subgap energies, it is reduced above the band edge.\label{fig:absorptance_bulk_pin}}
 \end{center}
\end{minipage}
 \end{figure}
 \begin{figure}[t!]
 \begin{minipage}{5.2cm}
 \begin{center}
\includegraphics[height=4.5cm]{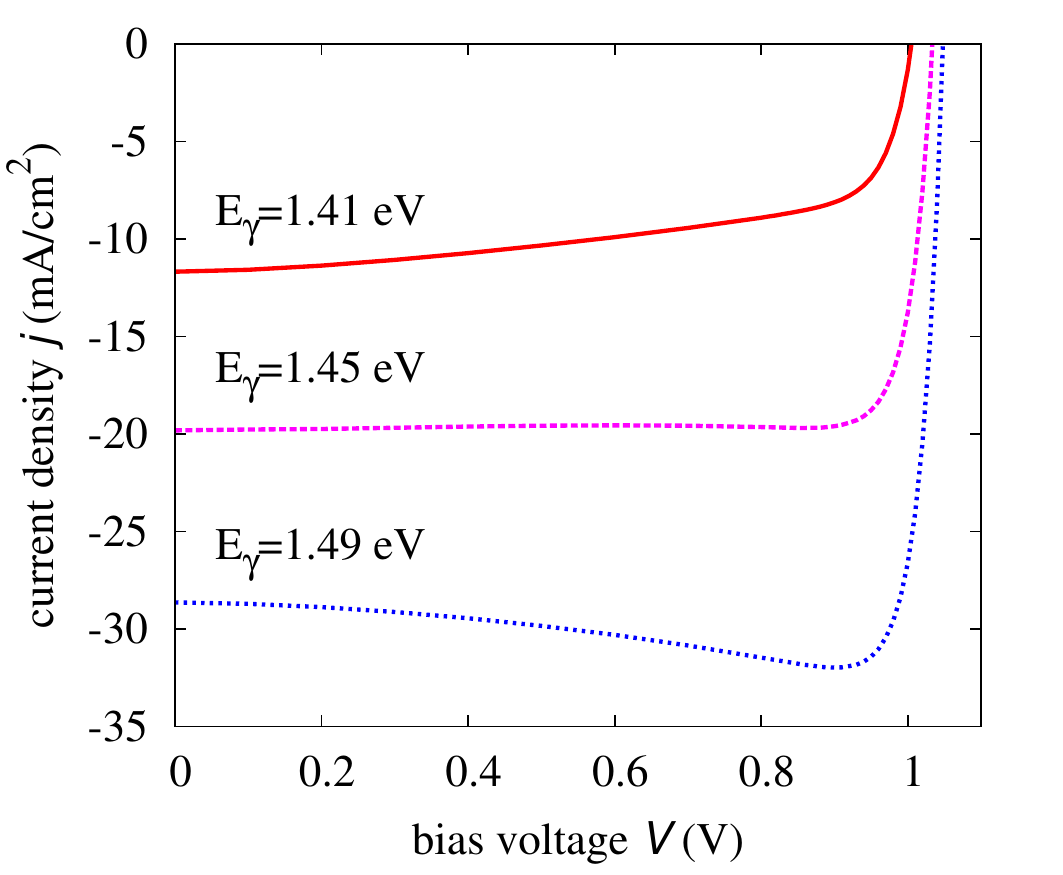}
 \caption{The bias dependence of the absorbance at different photon energies
 as displayed in Fig.~\ref{fig:absorptance_bulk_pin} is reflected in
 the monochromatic current-voltage characteristics (@0.1 kW/cm$^{2}$). \label{fig:jv_bulk_pin}}
 \end{center}
 \end{minipage}\quad
\begin{minipage}{5.2cm}
 \begin{center}
 \includegraphics[height=4.5cm]{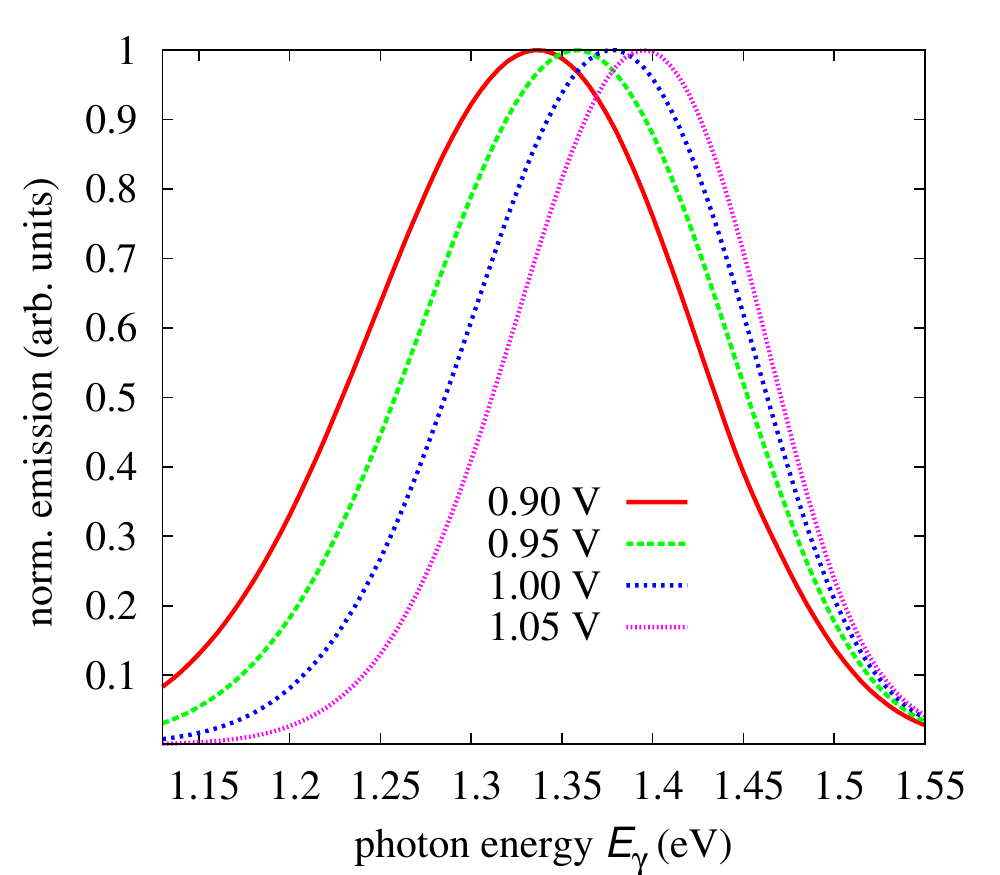}
 \caption{With the field-induced extension of the band edge to lower energies, the emission is also
 redshifted. \label{fig:emission_bulk_pin}\newline}
 \end{center}
\end{minipage}
\quad
 \begin{minipage}{5.2cm} 
 \begin{center}
\includegraphics[height=4.5cm]{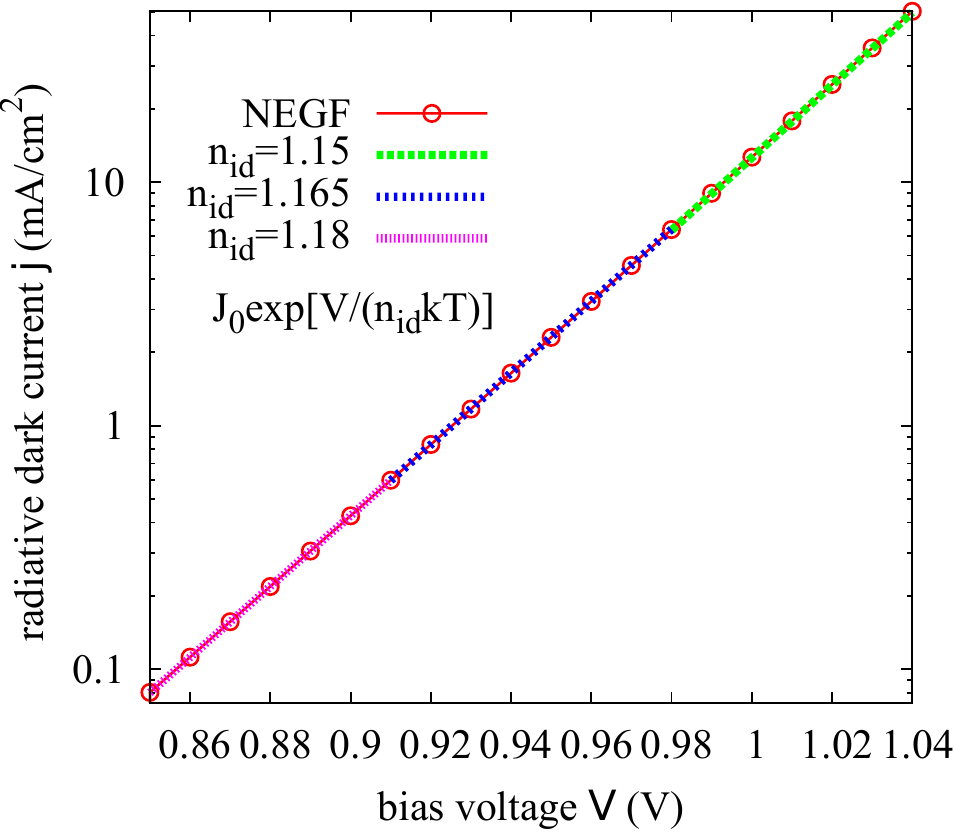}
 \caption{The characteristics of the radiative dark current show a
 tunnel enhancement of the ideality factor which decreases as the field
 is reduced.\label{fig:jv_dark_bulk_pin}}
 \end{center}
\end{minipage}
 \end{figure}
 
Recently, plasmonic concepts were presented for strong confinement of the incident light in
 absorber layers with thickness of only a few nm\cite{massiot:12,wang:13}. If such ultra-thin film
 solar cells remain based on bipolar junctions, such as the 40 nm GaAs $p-i-n$ diode with doping density of 3$\times 10^{18}$ cm$^{-3}$ investigated here using the EMA with parameters given in Tab.~\ref{tab:par_ema},  the doping-induced built-in potential is dropped
 over a very short distance (Fig.~\ref{fig:ldos_bulk_pin}), which results in strong doping- and bias-dependent band tailing (Franz-Keldysh) effects on the absorbance
 (Fig.~\ref{fig:absorptance_bulk_pin}), and consequently on the shape of the monochromatic current-voltage characteristics
 (Fig.~\ref{fig:jv_bulk_pin}).
 The bias dependent band tail leads not only to sub-gap absorption, but also to a red-shifting
 of the emission spectra under electrical injection at forward bias
 (Fig.~\ref{fig:emission_bulk_pin}). The decrease of the field with increasing bias causes the
 ideality factor of the radiative dark current to shrink towards unity as the field-induced 
 tunnel enhancement of the recombination process is reduced (Fig.~\ref{fig:jv_dark_bulk_pin}).

\subsection{Photocarrier transport in quantum well superlattice solar
cells}
\begin{figure}[b!]  
\begin{center}
%\begin{minipage}{10cm}
\includegraphics[height=6cm]{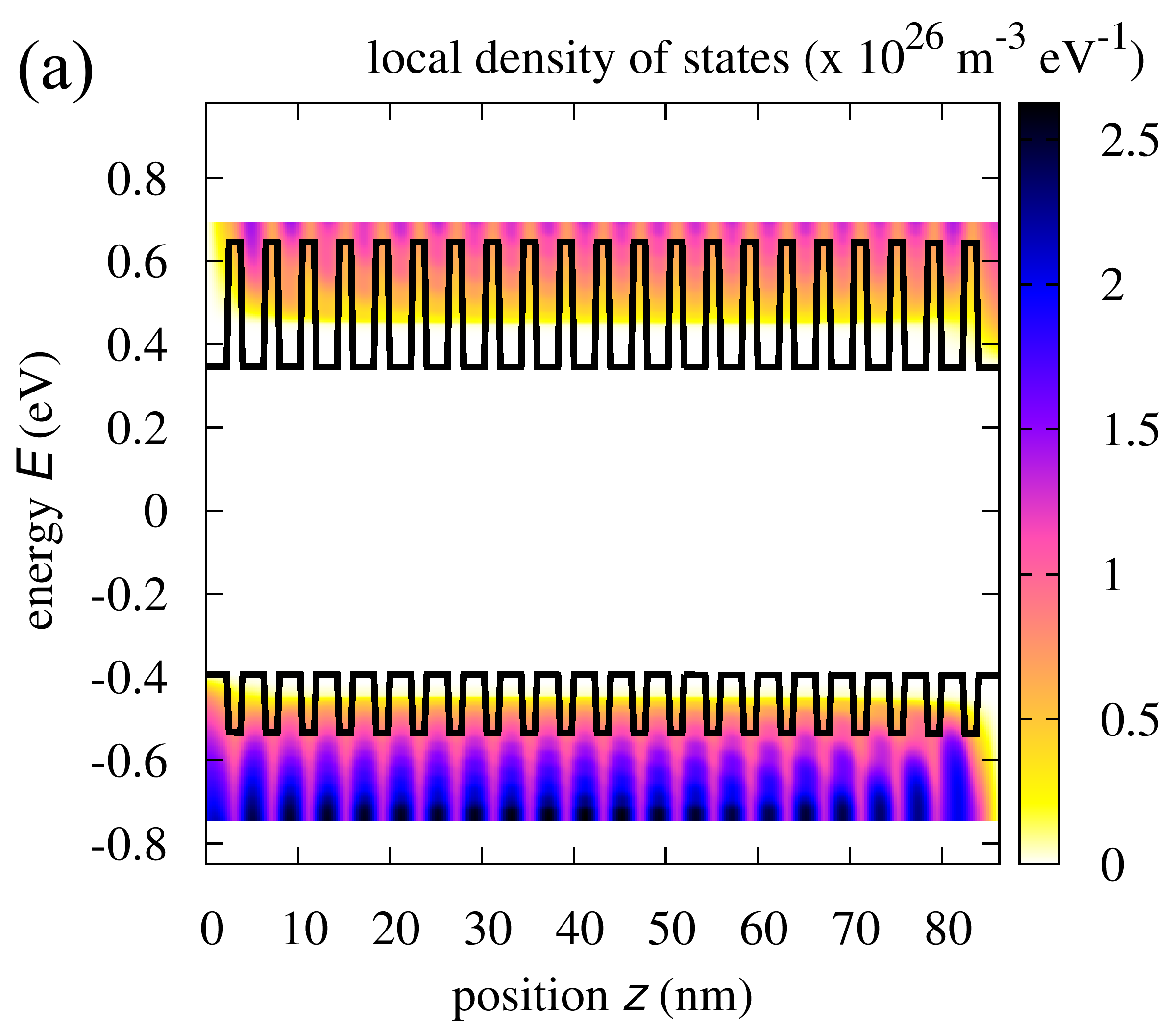}\quad\includegraphics[height=6cm]{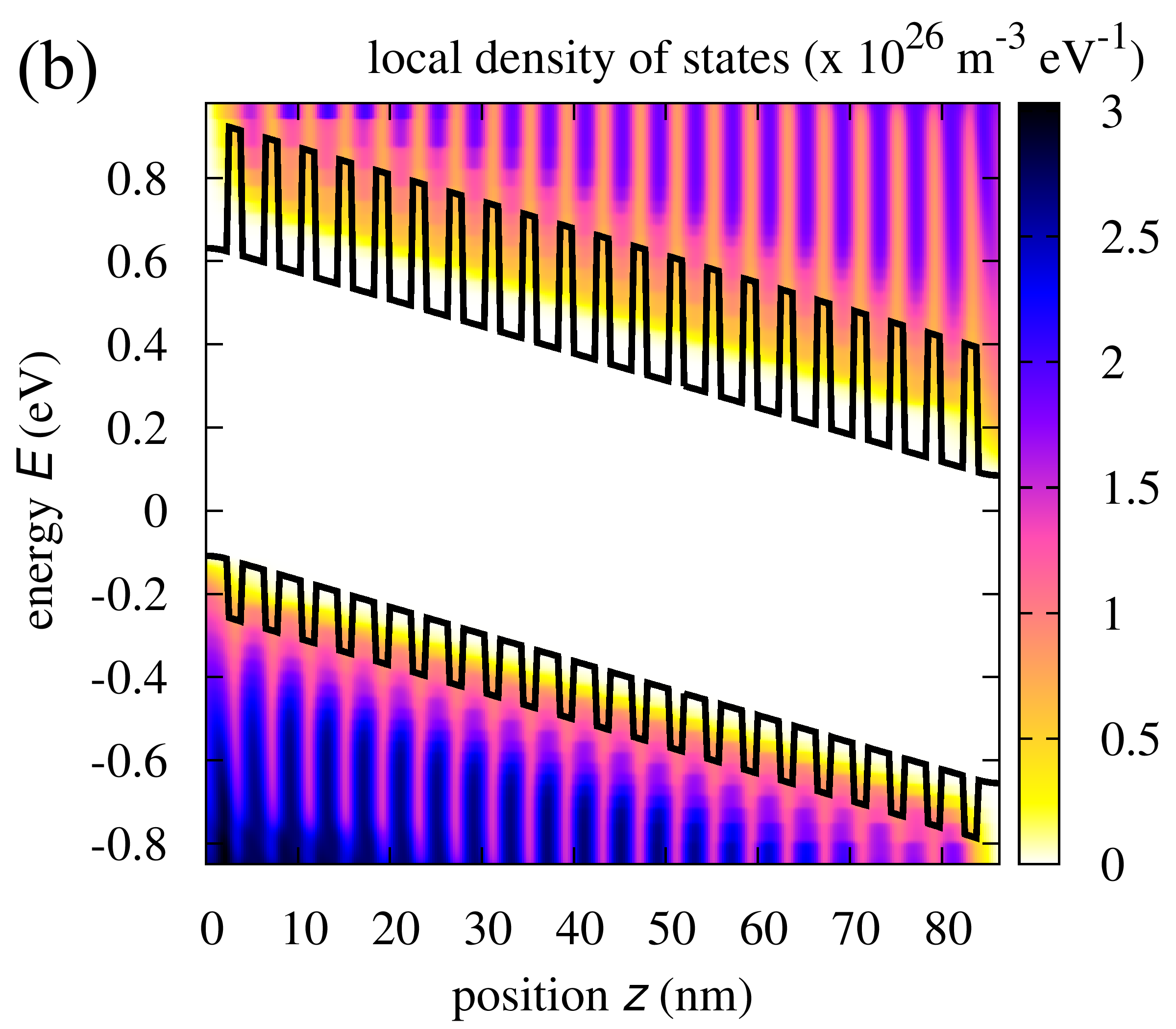}\caption{Local density of states of a 20 period quantum well superlattice at vanishing bias voltage and (a) flat band conditions, corresponding to an intrinsic system, and (b) for a doping density of $2\times 10^{18}$ cm$^{-3}$. While the effective band gap is increased as compared to bulk InGaAs due to confinement, the strong coupling results in a pronounced carrier delocalization and associated quasi-3d DOS.
 \label{fig:ldos_qwsl}}
 \end{center}
%\end{minipage}\quad
\end{figure}

Quantum well superlattice solar cells were proposed some time ago as tunable
band gap absorber components in multijunction solar cells \cite{green:00}. Recently, absorber
structures with strongly coupled quantum wells have shown enhanced carrier extraction efficiency as compared 
to multi quantum well structures with decoupled wells\cite{fujii:13}. Conventional approaches to the
simulation of quantum well superlattice solar cells assume either an infinite superlattice at flat
band conditions and band-like semiclassical transport\cite{kirchartz:09_SL} or ballistic transport
via the transfer matrix formalism \cite{courel:12}.

\begin{figure}[t!]
\begin{center}
\includegraphics[height=6.5cm]{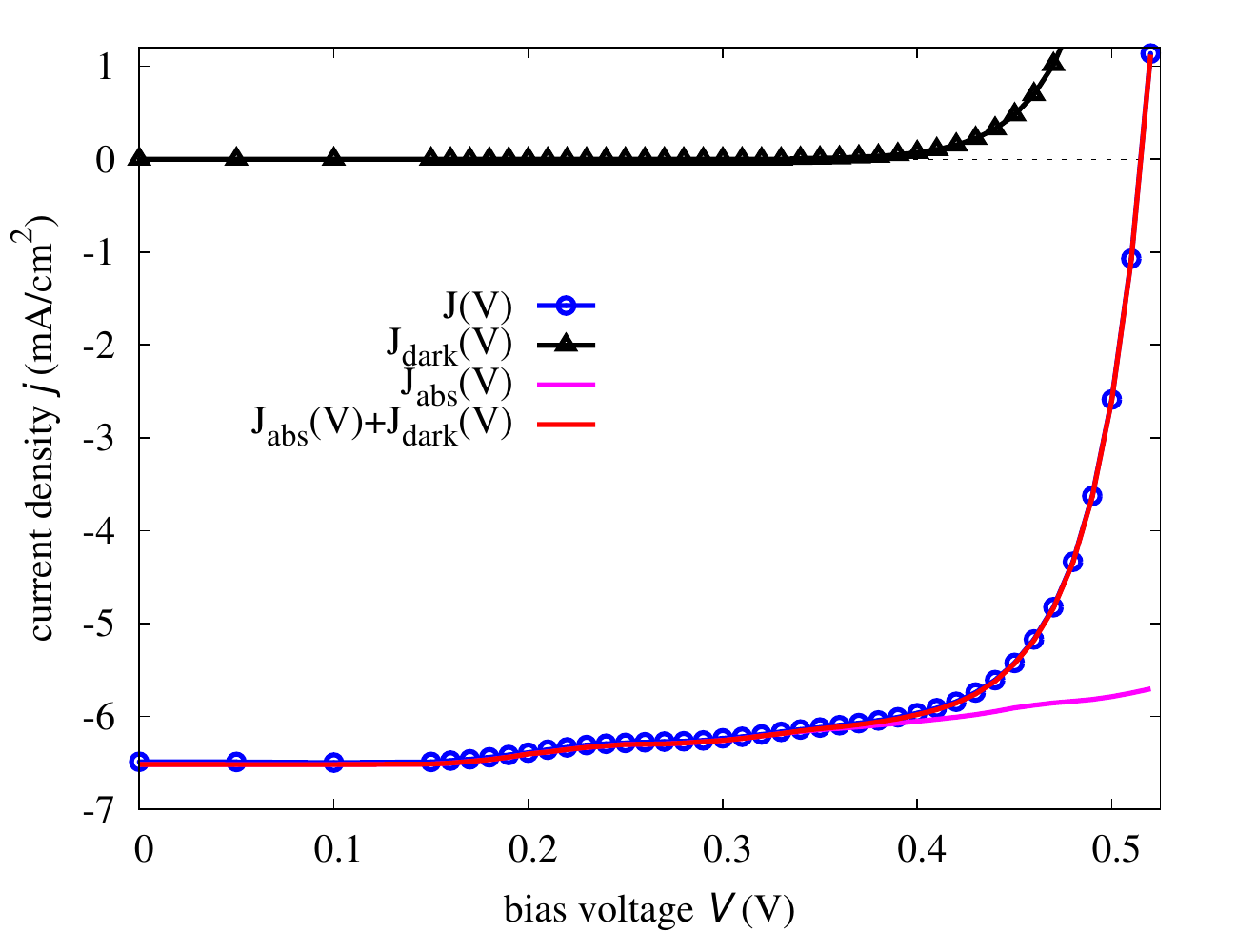}
\caption{Current voltage characteristics of the 20 period superlattice in the dark and under monochromatic illumination with photons of 0.95 eV and at 0.1 kW/cm$^{2}$. Due to the high degree of delocalization, the carrier escape probability is unity, i.e., the light JV-curve is the exact superposition of the bias dependent photocurrent from the absorptance and the radiative dark current.
 \label{fig:jv_qwsl}}
 %\end{minipage}
 \end{center}
\end{figure}    

\begin{figure}[t!]  
\begin{center}
\includegraphics[width=8cm]{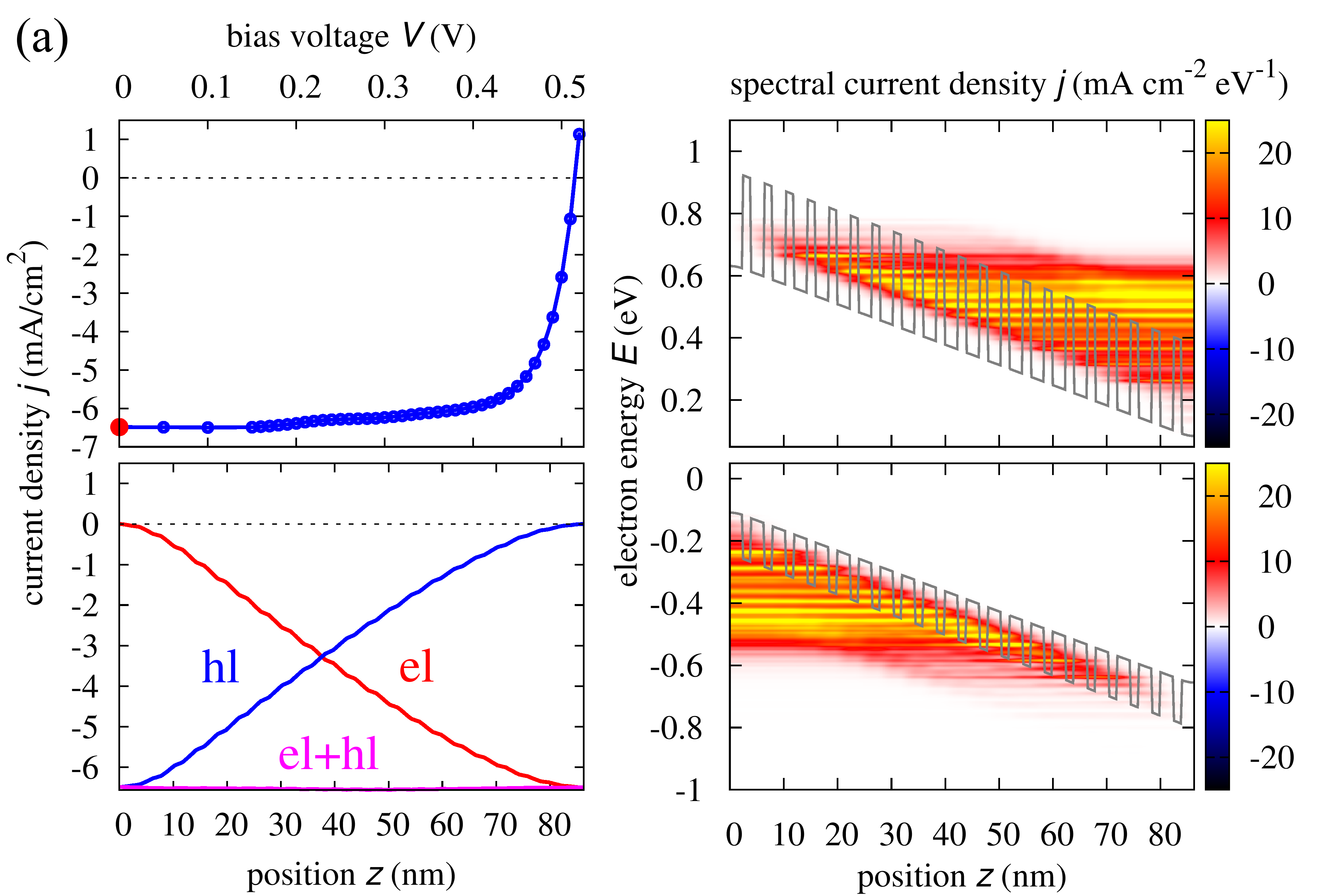}~\includegraphics[width=8cm]{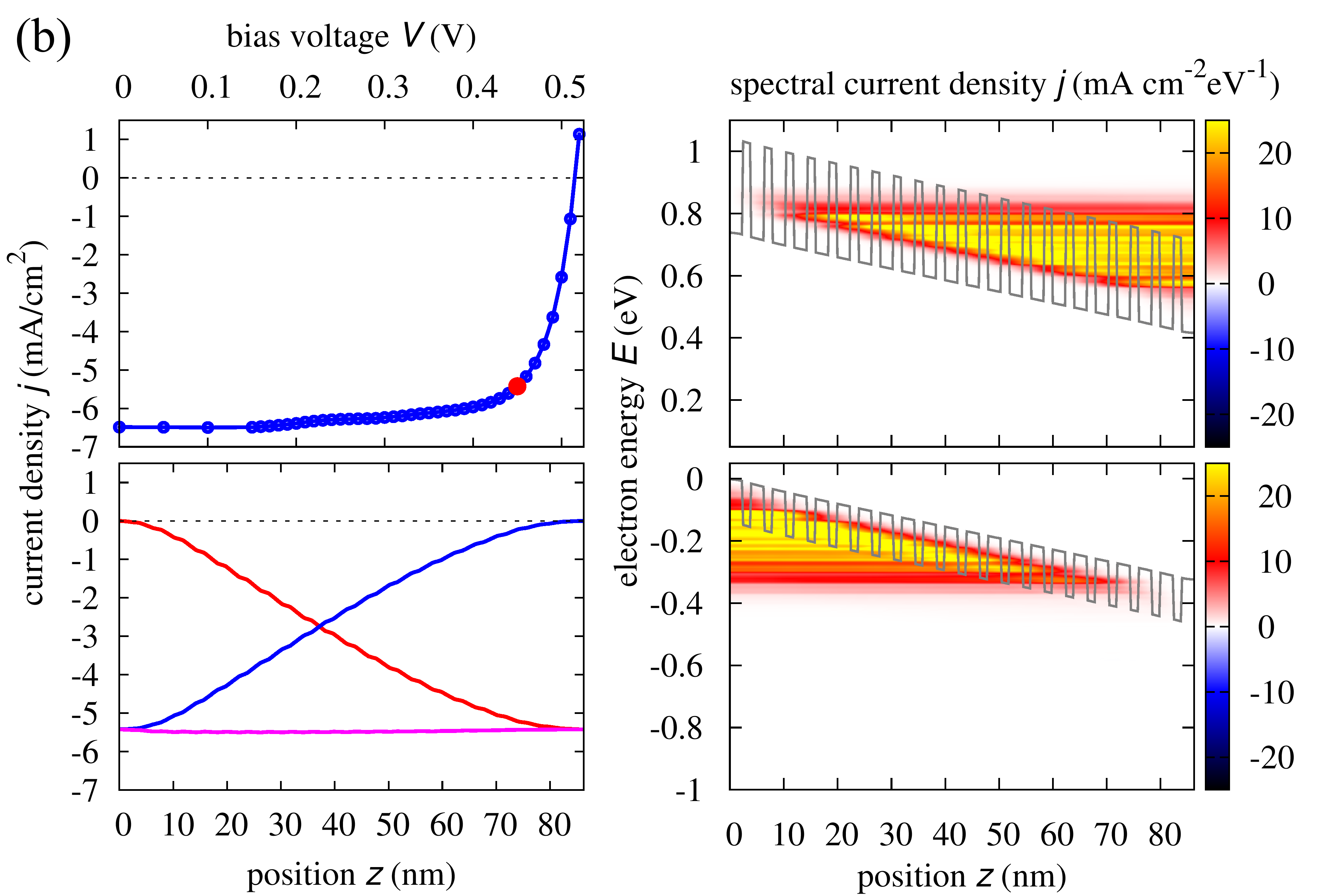}
\includegraphics[width=8cm]{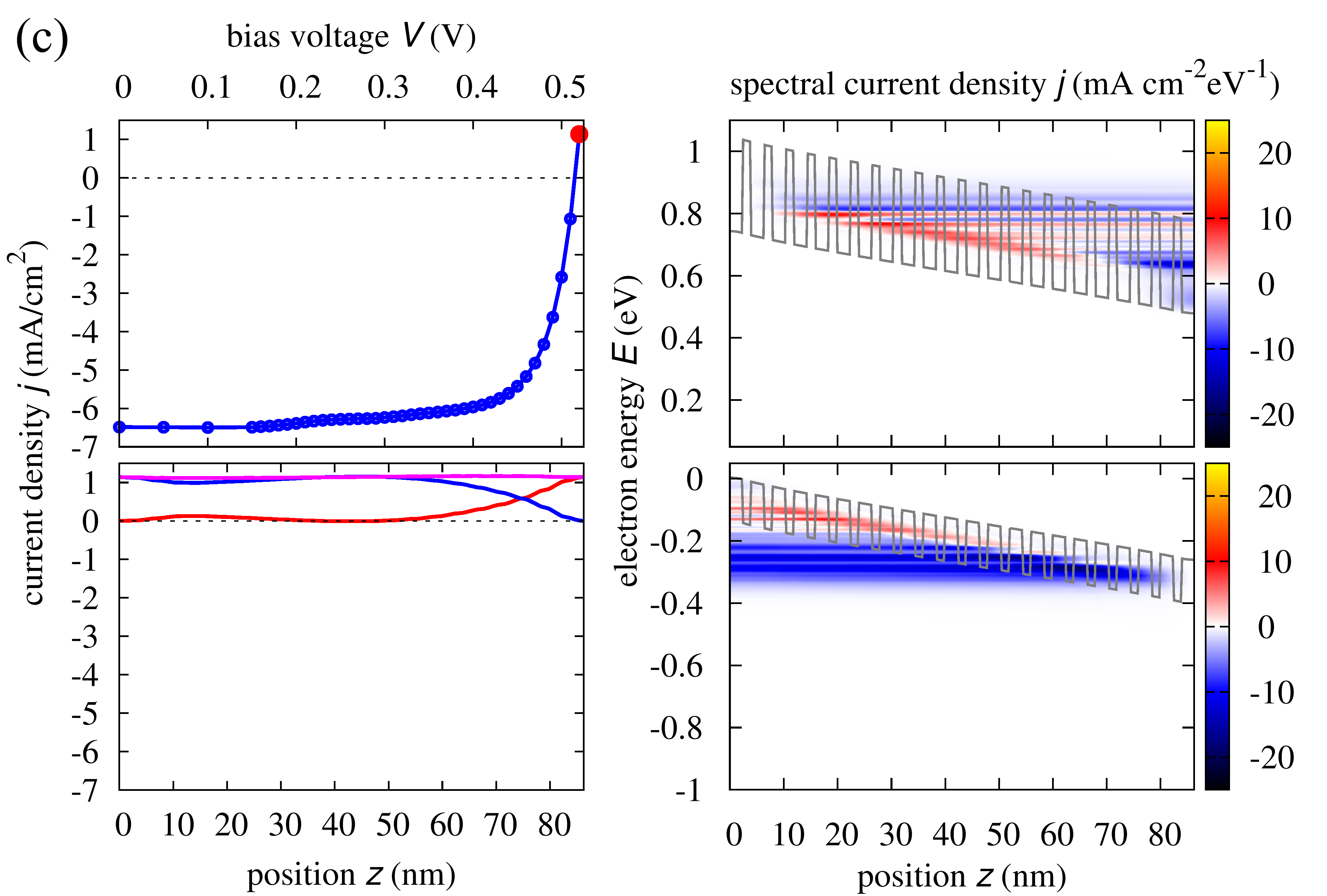}~\includegraphics[width=8cm]{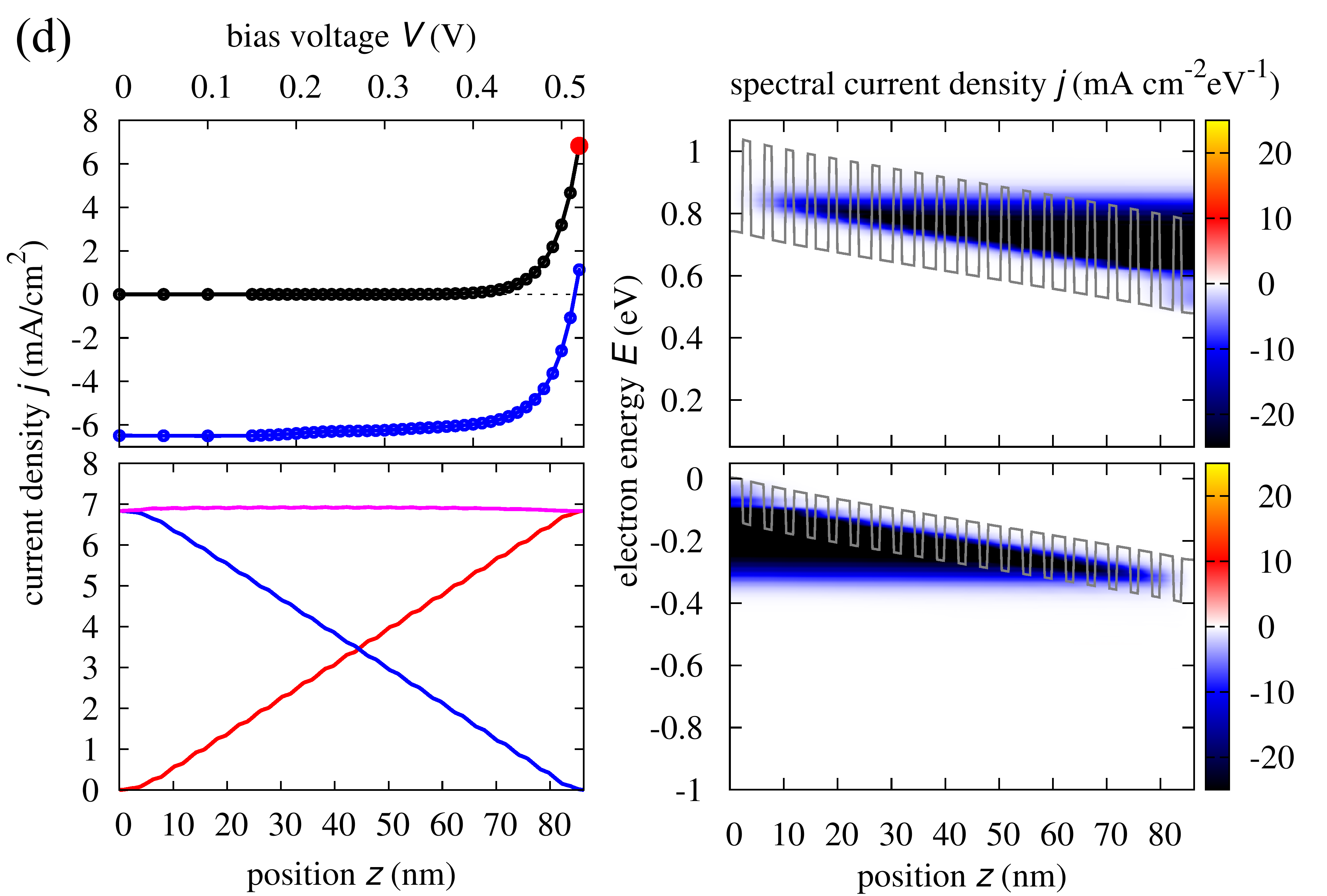}
\caption{JV characteristics, integral and spectral current flow at (a) short-circuit conditions (0 V), (b) close to the maximum power point (0.45 V), (c) at open-circuit conditions (0.52 V) and (d) at open-circuit voltage in the dark. While the minibands break up under realistic built-in fields, carrier extraction proceeds almost ballistically for thin barriers.
 \label{fig:currspect_qwsl}}
 \end{center}
\end{figure}

\begin{figure}[t!]  
\begin{center}
\includegraphics[width=16.2cm]{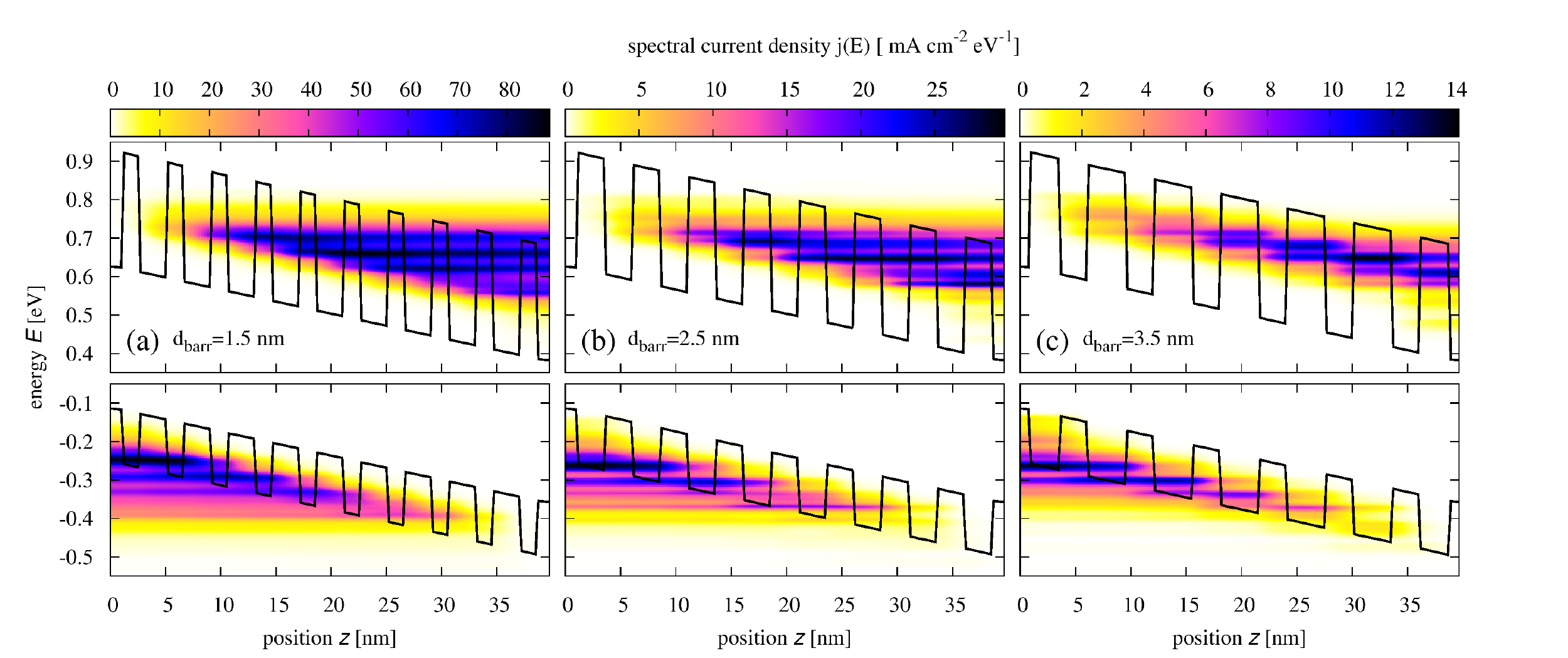}
 \caption{Photocurrent spectrum ($0.1$ kW/cm$^2$@0.95 eV) of coupled quantum well absorber
 structures with fixed InGaAs QW size of 2.5 nm and InAlGaAs layer thicknesses ranging from 1.5 nm
 to 3.5 nm. With increasing barrier width, the character of carrier transport gradually changes from quasi-ballistic to phonon-assisted sequential
 tunneling.
 \label{fig:spectcurrdens}}
 \end{center}
\end{figure}

Here, we use again the two-band EMA-NEGF approach. Figure \ref{fig:ldos_qwsl} shows the local density of states of a 20 period, selectively contacted In$_{0.52}$Al$_{0.33}$Ga$_{0.15}$As-In$_{0.53}$Ga$_{0.47}$As quantum well superlattice with barrier and well thicknesses of 1.5 nm and 2.5 nm, respectively, for (a) the flat band situation, corresponding to an intrinsic system, and (b) a $pin$-structure with strong built-in field induced by a doping density of $2\times 10^{18} cm^{-3}$. The strong coupling of the quantum wells results in a bulk-like density of states, however exhibiting an confinement induced increase in the effective band gap as compared to bulk InGaAs. Both features are preserved in the presence of the large built-in field. 

The (radiative) current-voltage characteristics of the 20 period superlattice structure is displayed in Fig.~\ref{fig:jv_qwsl} for applied forward bias voltage in the dark and under monochromatic illumination with photons of an energy of 0.95 eV and at an intensity of 0.1 kW/cm$^{2}$. The current-voltage characteristics $J(V)$ under illumination is perfectly reproduced by the superposition of the bias-dependent photocurrent $J_{abs}(V)$ as computed from the absorptance and the radiative dark current $J_{dark}(V)$. The charge carrier collection probability is therefore close to unity even at large bias voltages in the vicinity of the maximum power point.   

The charge carrier extraction can be investigated in more detail by considering the (spectral) current flow at different bias voltages. Fig.~\ref{fig:currspect_qwsl} shows the current-voltage characteristics together with spectral and integral current flow, for (a) short-circuit conditions (0 V), (b) close to the maximum power point (0.45 V), (c) at open-circuit conditions (0.52 V) and (d) at open-circuit voltage in the dark. The sum of integral electron and hole currents is conserved at all bias voltages. As can be inferred from the current spectra, the minibands break up under realistic built-in fields, but at strong coupling, i.e., for thin barriers, carrier extraction proceeds almost ballistically. However, as
shown in Fig.~\ref{fig:spectcurrdens} for structures of  coupled 2.5 nm wide InGaAs with InAlGaAs
barriers of thickness increasing from 1.5 nm to 3.5 nm, for thick barriers, the character of carrier transport changes from ballistic
to sequential tunneling assisted by phonon-mediated relaxation. Thus, a rigorous assessment of
carrier extraction in a given superlattice structure under realistic operating conditions is
provided at the quantum kinetic level only.

\subsection{Absorption losses in double quantum well tunnel junctions for multijunction solar cells}
In high-efficiency multijunction solar cells, double quantum well structures have been proposed as
tunnel junctions with high peak currents as well as low optical transmission losses
\cite{lumb:12}. In the junction region, the pronounced spatial variation of the doping profile gives
rise to extreme band bending effects. As shown by recent NEGF simulations employing a $sp_{z}$ tight-binding basis\cite{ae:prb87_13}, the
strong fields result in large deviations of the local density of states from the situation at flat band
conditions, apparent in both bulk injection and quantum well tunnel zones (Fig.~\ref{fig:ldos_TJ}a).
The local structure of bound and quasibound states in the junction region not only affects the
tunneling transport of charge carriers, but also the local absorption coefficient
(Fig.~\ref{fig:ldos_TJ}b). The associated absorbance no longer reflects the
joint density of states of either the bulk injection regions or a regular square well potential
(Fig.~\ref{fig:ldos_TJ}c).

\begin{figure}[b!]
\begin{center}
\includegraphics[height=5.2cm]{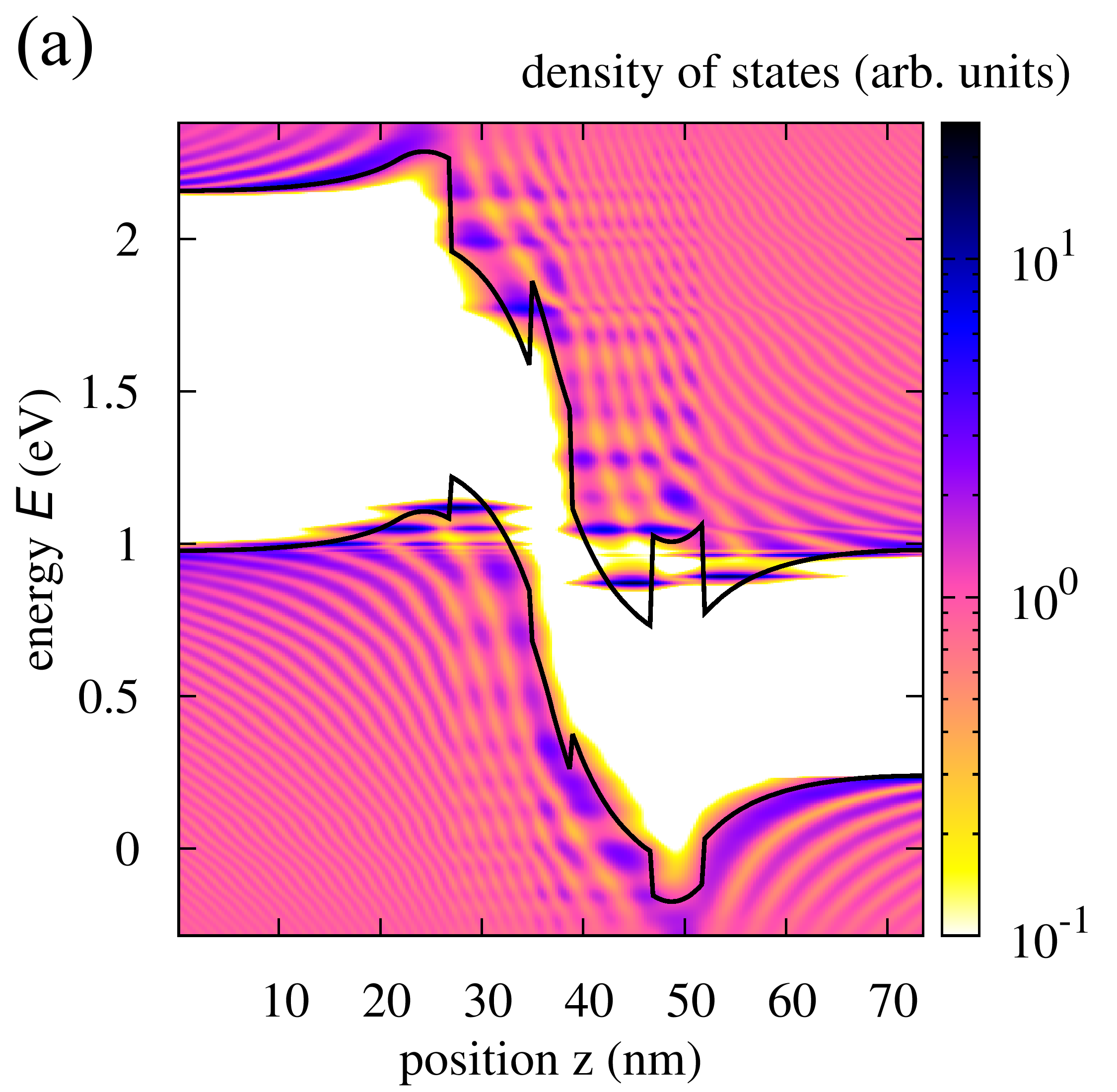}\includegraphics[height=5.2cm]{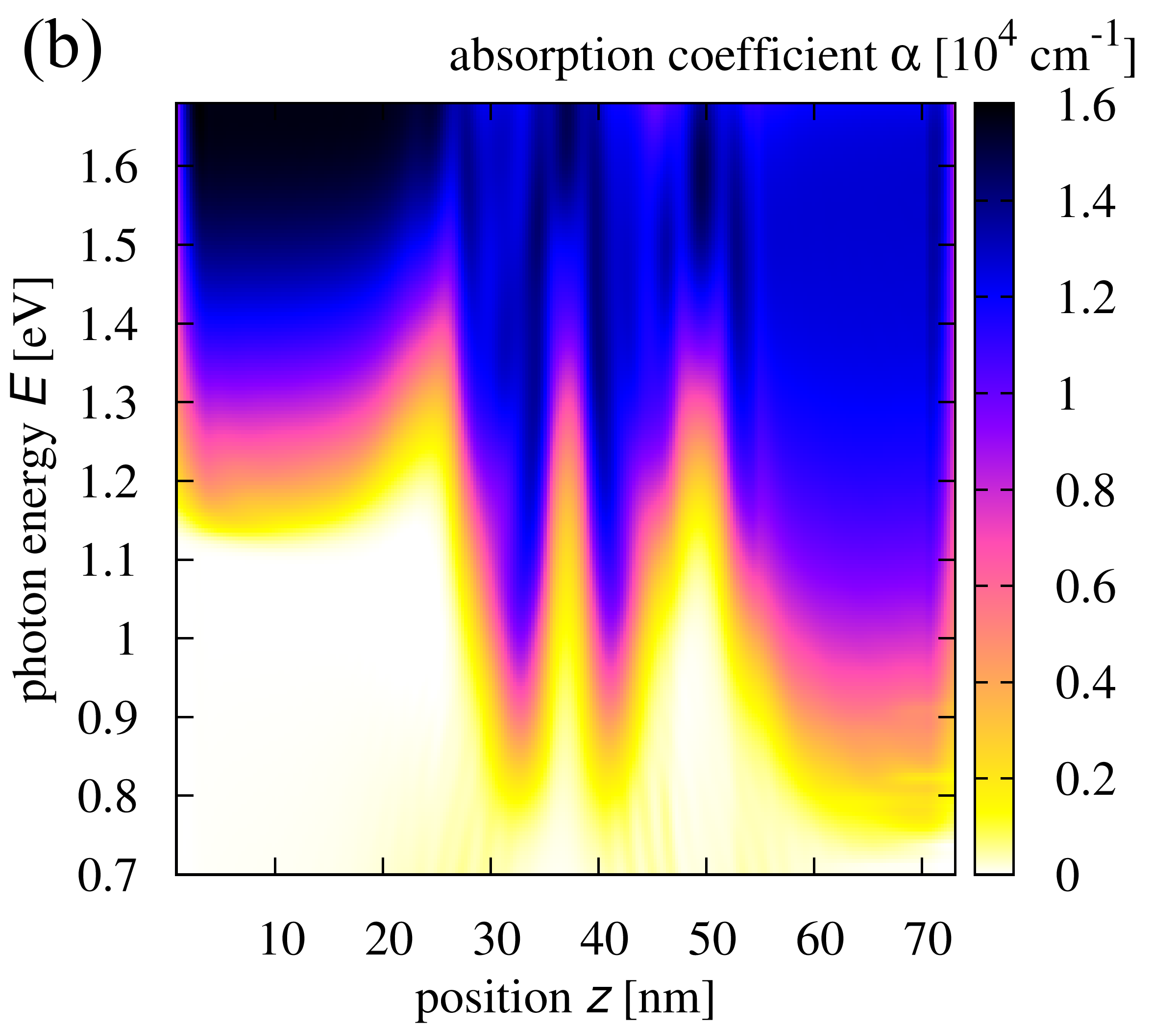}\includegraphics[height=4.9cm]{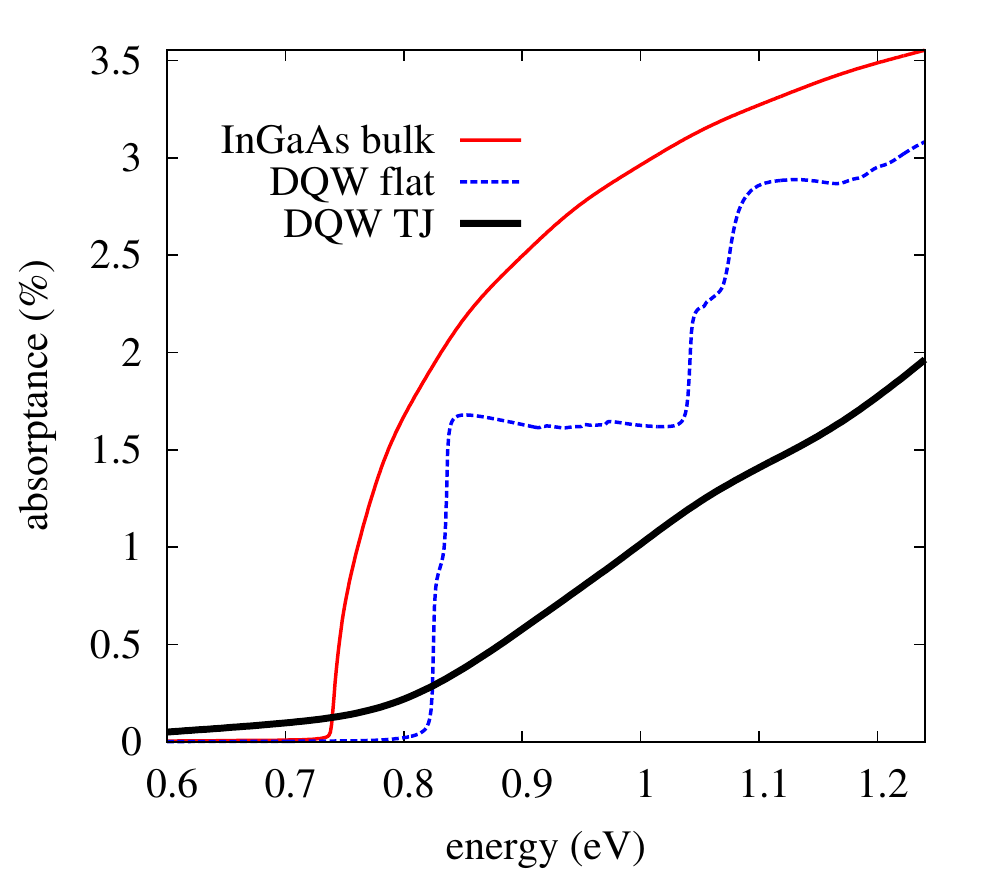}
\caption{(a) Local density of states in a double quantum well tunnel junction, (b) associated local
 absorption coefficient, and (c) the overall absorbance of the structure, revealing pronounced deviations
 from the flat band situation of a square well potential and bulk injection regions.
 \label{fig:ldos_TJ}} 
 \end{center}   
\end{figure}

\section{Conclusions} 
The strong local potential variations that are present in ultra-thin or nanostructure based solar
cell devices have a large impact on the photogeneration, photocarrier extraction and radiative
recombination processes, which is fully captured by the NEGF-based simulation framework proposed.
Notable examples are strong field induced band tailing effects in ultra-thin junction solar cells,
with associated red shift in the absorption and emission spectra. It is further shown that the
deviations from the flat band bulk picture are of special relevance for the assessment of carrier
transport in superlattice solar cells and of the absorption losses in novel tunnel junction architectures for
multijunction devices. 
 
\section{Appendix} 

 \begin{table}[h]
\caption{Material parameters for the two band effective mass model used for the simulation of the ultrathin film and quantum well superlattice absorbers.} 
\label{tab:par_ema}
{\small
\begin{center}       
\begin{tabular}{|l||r|r|r|l|} %% this creates two columns
%% |l|l| to left justify each column entry
%% |c|c| to center each column entry
%% use of \rule[]{}{} below opens up each row

\hline
\rule[-1ex]{0pt}{3.5ex}  & GaAs&In$_{0.53}$Ga$_{0.47}$As & In$_{0.52}$Al$_{0.33}$Ga$_{0.15}$As&\\
\hline\hline
\rule[-1ex]{0pt}{3.5ex}  $m^{*}_{el}/m_{0}$ &  0.063&0.041&0.065&electron effective mass\\
\hline
\rule[-1ex]{0pt}{3.5ex}   $m^{*}_{hl}/m_{0}$ &0.22 &0.052 &0.087&hole effective mass  \\
\hline
\rule[-1ex]{0pt}{3.5ex}  $E_{gap}$ (eV) & 1.42&0.74 & 1.18& band gap\\
\hline
\rule[-1ex]{0pt}{3.5ex}  $E_{aff}$ (eV)&4.07&4.54&4.24&electron affinity\\
\hline
\rule[-1ex]{0pt}{3.5ex}  $P_{cv}$ (eV)&28.8&18&18&optical matrix element\\
\hline
\rule[-1ex]{0pt}{3.5ex}  $\varepsilon_{r}$&13.18&14.24&13.18&relative dielectric constant\\
\hline
\rule[-1ex]{0pt}{3.5ex}  $\varepsilon_{\infty}$&10.89&8.16&10.72&high frequency dielectric constant\\
\hline
\rule[-1ex]{0pt}{3.5ex}  $D_{AC,el}$ (eV)&8.9322&8.9322&8.9322&acoustic deformation potential (el)\\
\hline
\rule[-1ex]{0pt}{3.5ex}  $D_{AC,hl}$ (eV)&5&5&5&acoustic deformation potential (hl) \\
\hline
\rule[-1ex]{0pt}{3.5ex}  $\rho$ (m$^{-3}$)&2329&2329&2329&material density\\
\hline
\rule[-1ex]{0pt}{3.5ex} $c_{s}$ (m~s$^{-1})$&9040&9040&9040&sound velocity\\
\hline
\rule[-1ex]{0pt}{3.5ex}  $\hbar\Omega_{LO}$ (eV)&0.036&0.033&0.038&optical phonon energy\\
\hline
\end{tabular}
\end{center}}
\end{table} 

\newpage

%%%%% References %%%%%
%\bibliography{spie14_paper_clean}   %>>>> bibliography data in report.bib
\bibliographystyle{spiejour}   %>>>> makes bibtex use spiejour.bst
%\input{spie14_paper_clean.bbl}

%%%%%%%%%%%%%%%%%%%%%%%%%%%%%%%%%%%%%%%%%%%%%%%%%%%%%%%%%%%%%
%%%%% Biographies of authors %%%%%

\vspace{2ex}\noindent{\bf Urs Aeberhard} is a senior researcher at the Institute of Energy and Climate Research 5:
Photovoltaics, Research Centre J\"ulich, Germany. He received the diploma degree  in theoretical physics and the PhD
degree in condensed matter theory from the Swiss Federal Institute of Technology in Zurich (ETHZ ) in 2004 and 2008, respectively. 
His current research interests include the theory and simulation of nanostructure based solar cell devices.

\listoffigures
%\listoftables

\end{spacing}
\end{document}